\documentclass[12pt,a4paper]{article}
\usepackage{times}
\usepackage{amssymb}
\usepackage{epsf}
\newlength{\HFPP}       \HFPP5.4mm
\makeatletter
\@addtoreset{equation}{section}
\makeatother

\def\preprint#1#2{\noindent\hbox{#1}\hfill\hbox{#2}\vskip 10pt}
\def\k_tilde{\sqrt{\frac{\kappa}{1-\kappa}}}
\def\kt2{\frac{\tilde{\kappa}}{2}}

\begin{document}
\begin{titlepage}
\def\thefootnote{\fnsymbol{footnote}}

\preprint{ITP-UH-2/97}{January 1997}
\vfill

\begin{center}
  {\Large\bf Properties of the chiral spin liquid state\\
  in generalized spin ladders}
\vfill

{\sc Holger Frahm}\footnote{e-mail: frahm@itp.uni-hannover.de}
	and 
{\sc Claus R\"odenbeck}\footnote{e-mail: roeden@itp.uni-hannover.de}
\vspace{1.0em}

{\sl
  Institut f\"ur Theoretische Physik, Universit\"at Hannover\\
  D-30167~Hannover, Germany}
\end{center}
\vfill

\begin{quote}
  We study zero temperature properties of a system of two coupled quantum
  spin chains subject to fields explicitly breaking time reversal symmetry
  and parity.  Suitable choice of the strength of these fields gives a
  model soluble by Bethe Ansatz methods which allows to determine the
  complete magnetic phase diagram of the system and the asymptotics of
  correlation functions from the finite size spectrum.  The chiral
  properties of the system for both the integrable and the nonintegrable
  case are studied using numerical techniques.
\end{quote}

{PACS-Nos.: 75.10.Jm, 75.30.Kz, 75.50.Ee}
\vspace*{\fill}
\setcounter{footnote}{0}
\end{titlepage}

%
%
\section{Introduction}
%
%
The idea of a chiral spin liquid state spontaneously breaking parity ($P$)
and time reversal ($T$) invariance has attracted considerable interest
recently.  It has first been proposed as a possible ground state for the
two dimensional $S={1\over2}$ Heisenberg model on a square lattice
frustrated with a sufficiently strong antiferromagnetic next-nearest
neighbour interaction \cite{wwz:89}.  Subsequent studies of this model have
found an enhancement of a chiral order parameter, comparison with other
possible states however suggests that the chiral spin state is unstable
\cite{Chiral4}.
Different lattices, in particular the triangular and Kagom\'e one, have
also been studied, however no firm evidence of a chiral spin state has been
found so far.  Here the frustration is a consequence of the lattice
geometry.

To characterize a chiral phase several `order parameters' have been
introduced: For lattices built from triangular plaquettes the vector
chirality \cite{Miya}
\begin{equation}
   \mathbf{X} = {\mathbf{S}}_1 \times {\mathbf{S}}_2
     + {\mathbf{S}}_2 \times {\mathbf{S}}_3
     + {\mathbf{S}}_3 \times {\mathbf{S}}_1
\end{equation}
with ${\mathbf{S}}_i$ being the three spins on the corners of a triangular
cell has been discussed.  While a spontaneous symmetry breaking in
$\mathbf{X}$ appears to be unlikely in an isotropic Heisenberg system its
properties have been studied in stacked triangular antiferromagnets with
\emph{XXZ} type anisotropy \cite{nish:93,male:95} which are realized in the
ABX$_3$-type compounds such as CsCuCl$_3$ (see articles in \cite{diep:b94}).

A rotationally invariant operator which has a non zero expectation value in
a phase with broken $P$ and $T$ symmetry is defined through \cite{wwz:89}
\begin{equation}
   \widehat{\chi}_{_{\langle 123\rangle}} = 
    {\mathbf S}_{1}\cdot({\mathbf S}_{2}\times{\mathbf S}_{3})\ .
\label{chi}
\end{equation}
Finally, a topological ``Cherns'' number measuring the dependence of a
quantum state on a twist in the boundary conditions has recently been
introduced by Haldane and Arovas and used to characterize the ground state
of a Heisenberg model on a hexagonal lattice subject to $PT$ breaking
fields \cite{haar:95}.  The actual computation of the Cherns number,
however, is restricted to rather small systems thus limiting its use in
studies of a phase diagram at present.

While the existence of a phase with spontaneous broken chirality in a
frustrated two-dimensional Heisenberg model has not been established so far
(a possible candidate may be the Kagom{\'e} lattice \cite{Lhu}) the experimental
studies of ABX$_3$ compounds indicate that the frustration leads to a rich
phase diagram if a magnetic field is applied
\cite{nish:93,minx:94,sssh:95}.

\begin{figure}
\begin{center}
\leavevmode
\epsfxsize=0.8\textwidth
\epsfbox{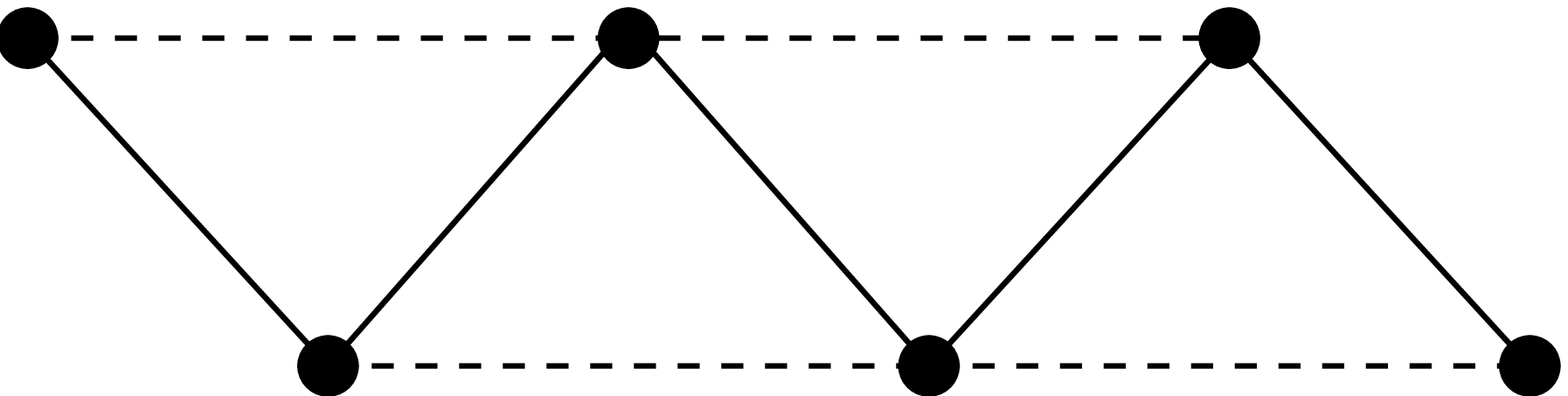}
\end{center}
\caption{
\label{fig:lattice} 
  Lattice on which the spin Hamiltonian (\ref{eq:Ham0}) is defined: The
  two-spin exchange coupling is $J_1$ and $J_2$ on full and dashed lines,
  respectively. The three spin exchange $\propto \chi_{1,2}$ couples the
  spins on the corners of each triangle.}
\end{figure}

Lacking a model with spontaneous broken $PT$-symmetry it is useful to
consider models containing terms breaking these symmetries
\emph{explicitly}.  Studies of such systems allow to gain a better
understanding of the properties of the chiral spin liquid state and means
for its characterization.  In this paper we analyze a system of two
spin-${1\over2}$ Heisenberg chains coupled by exchange terms on diagonal
bonds as shown in Fig.~\ref{fig:lattice} described by the Hamiltonian
\begin{equation}
  {\cal H}_0 = \sum_{n=1}^{2N} \left\{
	J_1 {\mathbf S}_n\cdot{\mathbf S}_{n+1} +
	J_2 {\mathbf S}_n\cdot{\mathbf S}_{n+2} \right\}
\label{eq:Ham0}
\end{equation}
with periodic boundary conditions ${\mathbf S}_{2N+k}={\mathbf S}_k$ and
even and odd indices labeling spins on the two sub chains respectively.
For $J_2=0$ this is just the Bethe Ansatz soluble Heisenberg chain
\cite{bethe:31}.  As long $J_2\le J_{2c}\approx0.25J_1$ the model continues
to have gapless excitations above a translationally invariant ground state
just as the single chain.  Increasing the frustrating next-nearest
neighbour interaction beyond $J_{2c}$ the system has two degenerate
dimerized ground states leading to the Majumdar--Ghosh model with a simple
dimer configuration for its ground states at $J_2={1\over2}J_1$
\cite{MaGh}.

To force the system into a chiral spin state we add terms breaking
$PT$-symmetry explicitly
\begin{equation}
  {\cal H}_\chi = \frac{1}{2}\sum_{n=1}^N\left\{
    \chi_1\ {\mathbf S}_{2n-1}\cdot
		({\mathbf S}_{2n}\times{\mathbf S}_{2n+1})
   -\chi_2\ {\mathbf S}_{2n}\cdot
		({\mathbf S}_{2n+1}\times{\mathbf S}_{2n+2})\right\}\ .
\label{eq:Hchi}
\end{equation}
Such multispin exchange terms may in fact be relevant for the description
of certain realizations of the two dimensional Heisenberg model such as
$^3$He layers on a graphite substrate \cite{godx:88}.  Note, that the
invariance under translations by one lattice site ($n\to n+1$) is destroyed
by these term unless $\chi_1=-\chi_2$.  For the full Hamiltonian including
the chiral terms one has additional integrable models:
for a ``staggered'' chiral field $\chi_1=-\chi_2$ the operator ${\cal
H}_\chi$ is one of the hierarchy of integrals of motion of the Heisenberg
chain, thus commutes with ${\cal H}_0(J_2=0)$.  The consequences of the
competition between these operators have been studied in \cite{frahm:92}.
Choosing the parameters in 
\begin{equation}
{\cal H}  = {\cal H}_0 + {\cal H}_\chi
\label{hami}  
\end{equation}
as
\begin{equation}
   J_1=2(1-\kappa)\ , \quad
   J_2=\kappa\  , \quad
   \chi_1=\chi_2=2\sqrt{\kappa(1-\kappa)}\
\label{parint}
\end{equation}
one obtains a family of integrable models of generalized spin ladders
introduced recently \cite{pozv:93,roeden:dipl,frcr:96}.  Varying of the
free parameter $\kappa$ from 0 to 1 the system evolves from a single
Heisenberg chain to a pair of decoupled ones, changing the sign of $\kappa$
reverses the sign of the chiral field, but doesn't affect most other
properties of the system.

Here we investigate the properties of the ground state and excitations of
this model.  In the following section the ground state energy and spectrum
of the low lying excitations of the integrable model (\ref{parint}) are
computed exactly.  As in the Heisenberg chain the excitations over the
antiferromagnetic (singlet) ground state are found to be spinons coming in
pairs.  For the characterization of the chiral properties we note that due
to its quasi one dimensional character it is not possible to define an
analogue of the topological Cherns number for this system.  In the
following we choose the expectation value of a uniform extension of the
chirality (\ref{chi}) as a measure of the chirality
\begin{equation}
   \widehat{\chi} = \frac{1}{N}\sum_{n=1}^N \left\{
    {\mathbf S}_{2n-1}\cdot({\mathbf S}_{2n}\times{\mathbf S}_{2n+1})
  - {\mathbf S}_{2n}\cdot({\mathbf S}_{2n+1}\times{\mathbf S}_{2n+2})
  \right\}\ .
\label{chiL}
\end{equation}
Unfortunately, expectation values of operators not commuting with the
Hamiltonian such as $\widehat{\chi}$ are not easily accessible within the
framework of the Bethe Ansatz.  Our results regarding
$\langle\widehat{\chi}\rangle$ are obtained from numerical diagonalization
of finite clusters.
In Section \ref{sec:intmag} we study the phase diagram of the integrable
model subject to a uniform external magnetic field ${\mathbf
h}\parallel\hat{z}$.  A characterization of the phases based on counting
the number of gapless excitations supported by the system is possible from
the Bethe Ansatz analysis.  We identify three different phases in the
$\kappa$--$h$ plane: for sufficiently large $h>h_{c2}$ the system shows
saturated ferromagnetism.  For ${1\over4}<\kappa<1$ a phase with low lying
excitations at four different wave numbers $\pm k_{1,2}$ is found for
magnetic fields $h_{c1}<h<h_{c2}$ (a similar phase diagram has recently
been established \cite{fufk:96} in an integrable chain of alternating spins
${1\over2}$ and $1$ \cite{AltSpin}).  To characterize these phases
we can compute the magnetization again from the Bethe Ansatz while we have
to rely on numerical results from finite systems for the chirality.
In the final section we summarize our findings and comment on the
properties of systems where the parameters $\chi_{1,2}$ are tuned away from
the integrable point.

\section{Ground state and excitations of the integrable model}
\label{sec:intgs}
As mentioned above, choosing the exchange constants for the Hamiltonian
(\ref{hami}) as in (\ref{parint}) gives a system which is integrable by
Bethe Ansatz methods.  Starting from the ferromagnetic state with all $2N$
spins pointing up one can reduce the solution of the Schr\"odinger equation
in the sector with $M$ overturned spins to a system of algebraic equations
\begin{equation}
  \left(\frac{\lambda_j+\kt2+\frac{i}{2}}{\lambda_j+\kt2-\frac{i}{2}}\right)^N
  \left(\frac{\lambda_j-\kt2+\frac{i}{2}}{\lambda_j-\kt2-\frac{i}{2}}\right)^N
  = \prod_{j\not= k}^M
	\frac{\lambda_j-\lambda_k+i}{\lambda_j-\lambda_k -i}\ ,
\qquad
  \tilde{\kappa} = \k_tilde 
\label{baeq}
\end{equation}
for the $M$ complex rapidities $\lambda_j$.  Each solution of these Bethe
Ansatz equations (\ref{baeq}) corresponds to an eigenstate of ${\cal H}$
with spin $S=N-M=S^z$.  Up to an overall constant the corresponding
eigenvalues are given by
\begin{equation}
  E(\{\lambda_j\}) = \sum_{j=1}^M \left(
	\tilde{\epsilon}_0(\lambda_j+\frac{\tilde{\kappa}}{2}) +
	\tilde{\epsilon}_0(\lambda_j-\frac{\tilde{\kappa}}{2})
	\right) + M h, \quad
  \tilde{\epsilon}_0(\lambda_j) = 
	-\frac{1}{2}\frac{1}{\lambda_j^2+\frac{1}{4}}\ .
\label{eps0}
\end{equation}
Here we have used the fact, that we are constructing eigenstates for fixed
$S^z$, to include the effect of an external magnetic field in the
Hamiltonian ${\cal H} \to {\cal H}- hS^z$.

A generic solution $\{\lambda_j\}$ of (\ref{baeq}) is organized groups of
uniformly spaced complex rapidities, so called \emph{strings}
\begin{equation}
   \lambda_j^{(m)} = x+i\mu_j\ ,\quad 
   \mu_j = -\frac{m+1}{2},-\frac{m+3}{2},\ldots,\frac{m-1}{2}.
\end{equation}
In the thermodynamic limit the ground-state is made up of real $\lambda$'s
only (1-strings).  Their density $\rho(\lambda)$ is given in terms of a
linear integral equation
\begin{equation}
 \rho(\lambda) + \int_{-\Lambda}^\Lambda d\mu K(\lambda-\mu) \rho(\mu) =
 \frac{1}{2\pi}\left(\frac{1}{(\lambda+\kt2)^2 + \frac{1}{4}} +
 \frac{1}{(\lambda-\kt2)^2 + \frac{1}{4}}\right)\ .
\label{str:rho}
\end{equation}
The kernel of this integral equation is $K(\lambda)=\frac{1}{2\pi}\,
\frac{2}{\lambda^2 +1}$.  The dependence on the magnetic field is
incorporated in the value of the integration boundaries $\Lambda$ which are
to be chosen such that the total density is $\int_{-\Lambda}^\Lambda
d\lambda\rho(\lambda)=(M/N)$.  For a vanishing external magnetic field
$h=0$ the ground state can be shown to be a singlet ($M=N$) and the
rapidities fill the entire real axis resulting in $\Lambda=\infty$ in
(\ref{str:rho}).  Hence $\rho$ can be computed by Fourier transform,
resulting in
\begin{equation}
   \rho(\lambda) = \tilde{\rho}(\lambda+\kt2) 
	+ \tilde{\rho}(\lambda-\kt2),\quad
	\mbox{with}\ \tilde{\rho}(\lambda) = \frac{1}{2\cosh(\pi\lambda)}.
\end{equation}
{}From (\ref{eps0}) the ground state energy $E_0$ per spin is given by
\cite{frcr:96}
\begin{eqnarray}
\frac{E_0}{2N} &=& \frac{1}{2}\int_{-\infty}^\infty d\lambda\rho(\lambda)
		   \epsilon_0(\lambda) 
		=  \int_{-\infty}^\infty d\lambda
  (\tilde{\rho}(\lambda+\kt2)\tilde{\epsilon}_0(\lambda) + 
  (\tilde{\rho}(\lambda-\kt2)\tilde{\epsilon}_0(\lambda) ) \nonumber \\
		&=& -\ln 2 -\frac{1}{4}(\Psi(1-\frac{i}{2}\tilde{\kappa}) - 
                        \Psi(\frac{1}{2}-\frac{i}{2}\tilde{\kappa}) +
			\Psi(1+\frac{i}{2}\tilde{\kappa}) -
			\Psi(\frac{1}{2}+\frac{i}{2}\tilde{\kappa})).
\end{eqnarray}
The term containing digamma--functions $\Psi(x)$ increases from $-\ln 2$  to 0
as $\kappa$ varies between 0 and 1. 

\begin{figure}
\begin{center}
\leavevmode
\epsfxsize=0.8\textwidth
\epsfbox{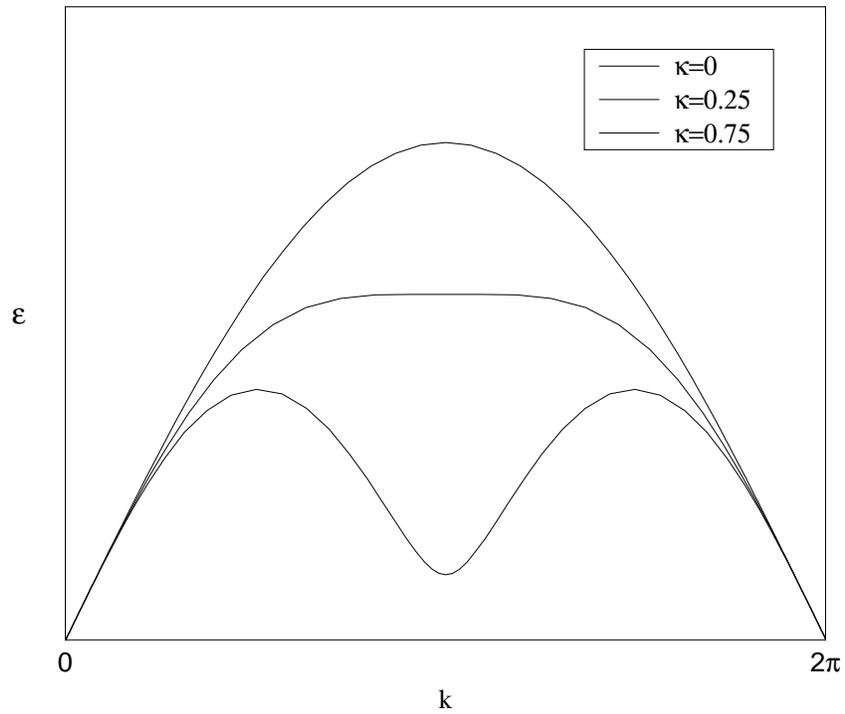}
\end{center}
\vspace*{.5cm}
\caption{
\label{fig:disp} 
   Spinon dispersion of the integrable model (\ref{parint}) for several
   values of $\kappa$.}
\end{figure}

Low lying excitations of the system are parametrized by holes in the
distribution of real $\{\lambda_j\}$. The dispersion of these
\emph{spinons} is determined by the integral equation for the {\em dressed
energies}:
\begin{equation}
 \epsilon(\lambda) + \int_{\epsilon(\lambda)<0}
	d\mu K(\lambda-\mu) \epsilon(\mu)  =  h
   -\frac{1}{2}\left(\frac{1}{(\lambda+\kt2)^2 + 
    \frac{1}{4}} + \frac{1}{(\lambda-\kt2)^2 + \frac{1}{4}}\right).
\label{str:epsilon}
\end{equation}
In this grand canonical approach the ground state is characterized as the
one in with all states with negative dressed energy $\epsilon(\lambda)$ are
filled.  For vanishing or sufficiently small (see below) magnetic fields
this condition is related to that used in (\ref{str:rho}) above through
$\epsilon(\pm\Lambda)=0$.  Again, we can solve (\ref{str:epsilon}) by
Fourier transform for $h=0$, giving
\begin{eqnarray}
  \epsilon(\lambda)=
     \tilde{\epsilon}(\lambda+\frac{\tilde{\kappa}}{2}) +
     \tilde{\epsilon}(\lambda-\frac{\tilde{\kappa}}{2})\ ,&& \quad
  \tilde{\epsilon}(\lambda) = \frac{\pi}{2\cosh(\lambda)} 
\nonumber \\
  k(\lambda) = \tilde{k}(\lambda+\frac{\tilde{\kappa}}{2}) +
     \tilde{k}(\lambda-\frac{\tilde{\kappa}}{2})\ , && \quad
     \tilde{k}(\lambda) = \arctan(\lambda) - \frac{\pi}{2}\ . 
\label{dreEn}
\end{eqnarray}
Eliminating $\lambda$ from these equations one obtains the spinon
dispersion $\epsilon(k)$ which is shown for several values of $\kappa$ in
Figure \ref{fig:disp}.

Comparing these results with the known solution of the usual $S={1\over2}$
XXX Heisenberg chain \cite{fata:84} we find that they coincide in the
limits $\kappa\to 0,1$ as is expected from the Hamiltonian.  Zero
temperature quantities such as densities of rapidities or their dressed
energies are simply superpositions of the corresponding quantities for the
XXX chain with argument shifted by $\tilde{\kappa}$ (see (\ref{str:rho}),
(\ref{str:epsilon})).  On the basis of the properties of the low lying
excitations this comparison can be extended to the critical properties of
the system: For zero $h$ the continuum limit of the system can be
identified as a conformal field theory (CFT) with central charge $c=1$ for
any $0\le\kappa<1$.  As for the XXX chain we expect this situation to be
described by a level-1 SU(2) Wess-Zumino-Witten model.  For $\kappa=1$
there appears another massless mode leading to low energy properties
corresponding to two $c=1$ models (see also Section~\ref{sec:intmag} below).

Finally we have studied the effect of the symmetry breaking terms ${\cal
H}_\chi$ in the integrable Hamiltonian.  It is instructive to start with
the exact solution of a system of four spins (see also \cite{wwz:89}).  The
Hamiltonian is given by (\ref{eq:Ham0}) with $N=2$ and the couplings are
\begin{equation}
   J_1 = 1, \quad J_2, \quad \chi_1=\chi_2=:\kappa, \quad 0\leq\kappa\leq 1.
   \label{choice}
\end{equation}
{}From the exact solution we know that the ground state of the integrable
model is always a singlet.  The same holds for the model given by
(\ref{choice}).  Hence it is sufficient to diagonalize the Hamiltonian in
the two-dimensional singlet subspace.  Depending on $J_2$ there are two
distinct cases:

\noindent
1. $J_2=\frac{1}{2}: \quad$ At the Majumdar--Ghosh point the singlets are
degenerate for vanishing chiral field whereas for finite $\kappa$ the
operator $\widehat{\chi}$ (\ref{chiL}) lifts this degeneracy.  One also
observes that the Hamiltonian simplifies to ${\cal H}={\mathbf S}^2 +
\kappa\widehat{\chi}$ (up to a constant) and $[{\mathbf S}^2,
\widehat{\chi}] = 0$.  Therefore the chirality $\widehat{\chi}$ can be
diagonalized in the singlet subspace, yielding a $\kappa$-independent
chirality which is found to be one of its eigenvalues
$\langle\widehat{\chi}\rangle = \pm \frac{1}{2}\sqrt{3}$.

\noindent
2. $J_2 \neq \frac{1}{2}: \quad$ Tuning $J_2$ away from $\frac{1}{2}$ the
ground state of the $\kappa=0$-system is unique.  At this point the
Hamiltonian is $P$-- and $T$--symmetric leading to a vanishing chirality in
the ground state.  Switching on the chiral field the two singlet
eigenstates of the Hamiltonian have non-zero expectation values
$\langle\widehat{\chi}\rangle$:
\begin{equation}
  \langle\widehat{\chi}\rangle = \pm {{3}\over2}\ 
	{\kappa \over \sqrt{3\kappa^2 + 2(2J_2-{1})^2}}\ .
\end{equation}
The ``chiral susceptibility'' $\partial\langle\widehat{\chi}\rangle/
\partial\kappa$ diverges for $\kappa=0$ at the MG point.  For $\kappa\gg1$
the chirality approaches its eigenvalue $\frac{1}{2}\sqrt{3}$.

\begin{figure}
\begin{center}
\leavevmode
\epsfxsize=0.8\textwidth
\epsfbox{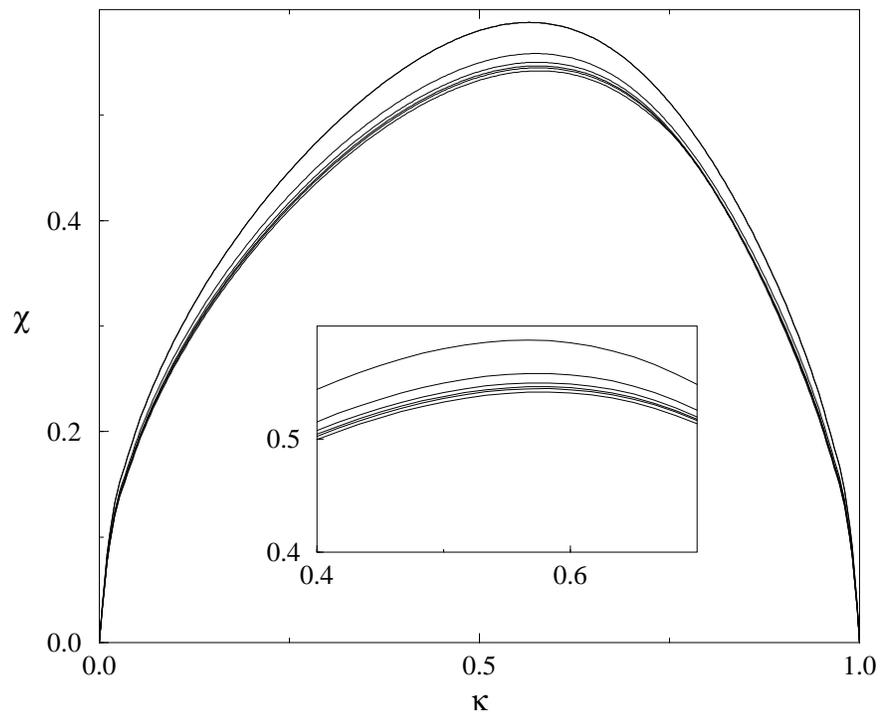}
\end{center}
\vspace*{.5cm}
\caption{
\label{fig:chi} 
  Chirality of the integrable Model (\ref{parint}). The system sizes are
  $2N=8,12,16,20,24, \infty$ from top to bottom.  }
\end{figure}

For larger systems no exact results on the chirality can be obtained (as
mentioned in the introduction it is not possible to compute this
expectation value directly from the Bethe Ansatz solution).  For small
systems (up to 24 spins) we have used a Lanzcos algorithm to compute some
low lying states numerically.  In Fig.~\ref{fig:chi} we present our results
on ground state chirality from these data: $\langle\hat{\chi}\rangle$
vanishes for $\kappa\to0,1$ as is expected from the single chain
Hamiltonian in these limits.  For intermediate values of $\kappa$ we find
that the finite size corrections to the chirality are very small, so that
the properties of the infinite system can easily be read off from the
numerical data.  We find that the maximum of the chirality is obtained at
$\kappa\approx0.58$ with about two thirds of the largest possible value
$\sqrt{3}/2$.
\section{Magnetic phase diagram of the integrable model}
\label{sec:intmag}
\begin{figure}
\begin{center}
\leavevmode
\epsfxsize=0.8\textwidth
\epsfbox{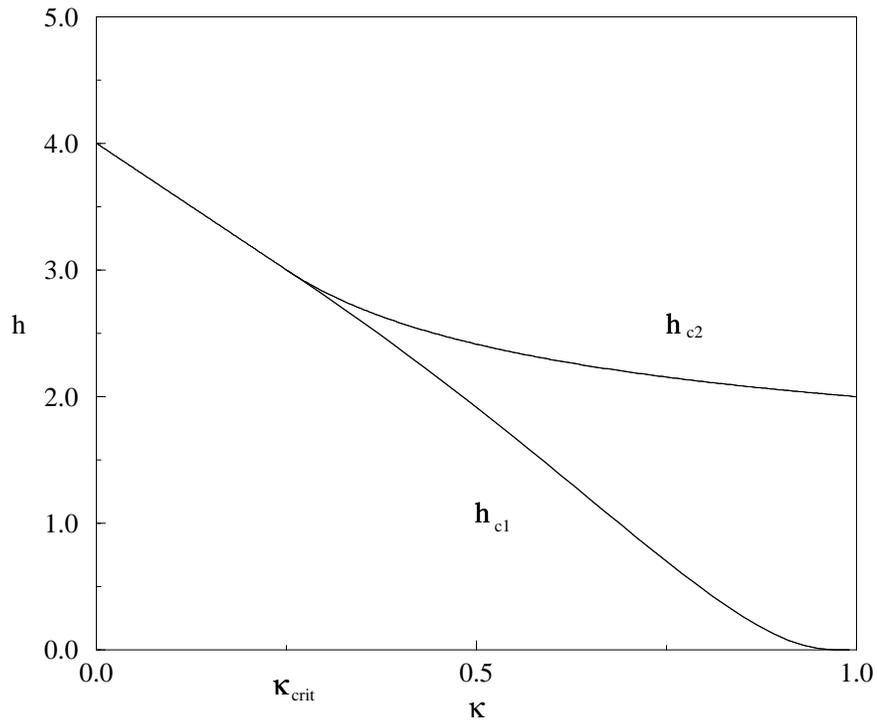}
\end{center}
\vspace*{.5cm}
\caption{
\label{fig:crit} 
  Magnetic phase diagram of the integrable system.  Below $h_{c1}$ the
  system has central charge $c=1$, for $h_{c1}<h<h_{c2}$ the model is a
  realization of two $c=1$ models, and for $h>h_{c2}$ the system saturates
  in a fully polarized (ferromagnetic) state.}
\end{figure}

While the zero field properties of the model (\ref{hami}) are closely
related to those of the single XXX Heisenberg chain, addition of an
magnetic field gives rise to an interesting phase diagram (see
Fig.~\ref{fig:crit}): For small values of $\kappa$ we find that the
magnetic field first breaks the SU(2) symmetry giving a $c=1$ Gaussian CFT
with anomalous dimensions depending on the magnetic field.  Increasing the
magnetic field $h$ beyond a value $h_{c1}(\kappa)$ two additional ``Fermi
points'' with gapless excitations arise, leading to a low energy spectrum
as described by two $c=1$ Gaussian CFTs.  Finally for magnetic field
$h>h_{c2}$ the ground state of the system saturates ferromagnetically.

\begin{figure}
\begin{center}
\leavevmode
\epsfxsize=0.8\textwidth
\epsfbox{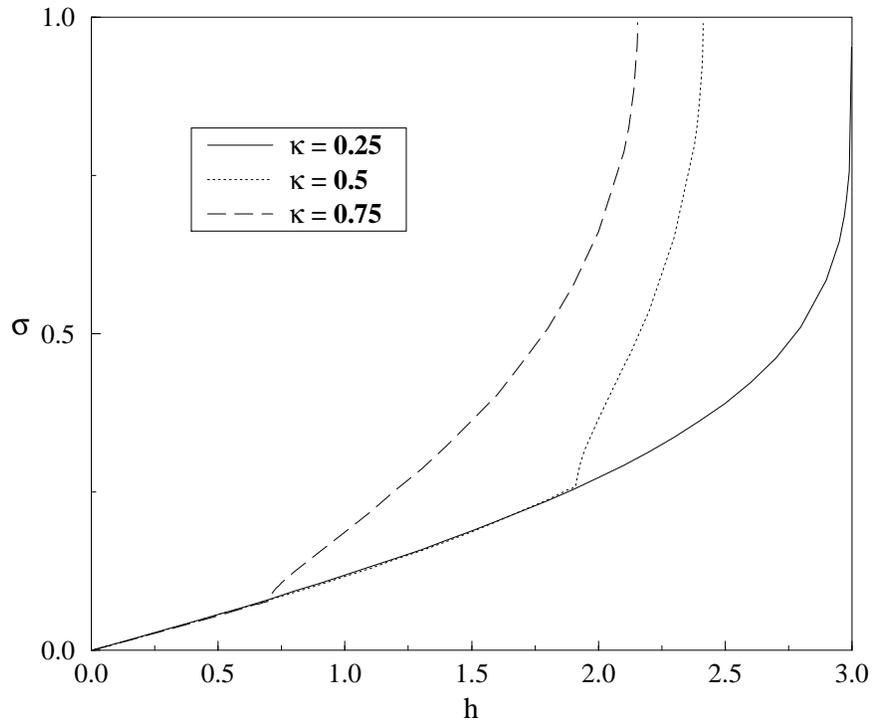}
\end{center}
\vspace*{.5cm}
\caption{
\label{fig:mag} 
  Magnetisation of the integrable model vs.\ magnetic field for three
  values of $\kappa$.}
\end{figure}

To calculate the corresponding phase boundaries as a function of $\kappa$
and the expectation values of the magnetization $\sigma=\langle
S^z\rangle/(2N)$ as a function of the magnetic field we have to solve the
integral equation (\ref{str:epsilon}).  The numerical results for the
magnetization are given in Figure \ref{fig:mag}.  

\begin{figure}
\begin{center}
\leavevmode
\epsfxsize=0.8\textwidth
\epsfbox{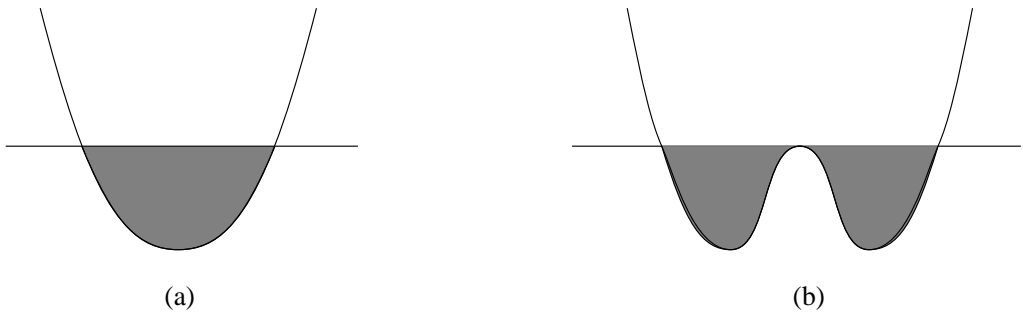}
\end{center}
\vspace*{.5cm}
\caption{
\label{fig:phases} 
  Filling fo the Fermi--Sea in presence of a magnetic field. Fig. (a):
  $\kappa < 1/4$ and $0<h<h_{c2}$.
  Fig. (b): $\kappa > 1/4$ and $h=h_{c1}$, the position of the 'kink'
  in Figure \ref{fig:mag}.} 
\end{figure}

Analytical results can be obtained near the critical field $h_{c2}$ which
is determined from the condition that $\epsilon(\lambda)\ge 0$ for all
values of $\lambda$.  For fields $h\lessapprox h_{c2}$ the integral
equation can be solved by iteration which allows to study the nature of low
lying excitations and the dependence of the magnetization on $h$.  Here one
has to distinguish three different cases:

(i) For $0\le\kappa<{1\over4}$ the minimum of the bare dispersion given by
the driving terms in (\ref{str:epsilon}) is at $\lambda=0$ (see
Fig.~\ref{fig:phases}(a)).  The dressed energies are nonnegative for
magnetic fields
\begin{equation}
   h \ge h_{c2} = 4(1-\kappa)\ .
\end{equation}
In this region the magnetization reaches its maximum according to a square
root law
\begin{equation}
  \sigma = 1-\frac{2}{\pi}\sqrt{\frac{1}{h_{c2}-3}}\sqrt{h_{c2}-h}\ .
\end{equation}
The system has massless excitations near the pseudo Fermi points
$\epsilon(\pm\Lambda)=0$.  Letting $h\to0$ one finds that there is no
additional phase transition, hence we can identify these excitations with
the spinons forming the excitation spectrum at zero magnetic field.
Standard techniques \cite{vewo:85} can be used to compute the spectra of
finite size systems from the Bethe Ansatz equations (\ref{baeq}).  The low
lying energies compared to the ground state energy of the infinite system
are ($L\equiv 2N$)
\begin{equation}
  \Delta E = -{\pi\over 6L}v 
  + {2\pi\over L}v\left(\Delta^{+}+\Delta^{-}\right)
  \label{fse1}
\end{equation}
where $v$ is the velocity of the massless magnons and the conformal dimensions
of primary operators are
\begin{equation}
  \Delta^{\pm} = {1\over2} \left\{ 
    {1\over2\xi}\Delta M \pm \xi\Delta D\right\}^2\ .
  \label{dim1}
\end{equation}
$\Delta M$ is an integer denoting the change in $S^z$ induced by the
operator, $\Delta D$ is an integer or half integer proportional to the
momentum of the excited state (due to backscattering).  The dressed charge
$\xi=\xi(\Lambda)$ is given in terms of the linear integral equation
$\xi(\lambda) + \int_{-\Lambda}^{\Lambda} d\mu K(\lambda-\mu) \xi(\mu) = 1$.
Depending on the external magnetic field it varies between $1/\sqrt{2}$ for
$h=0$ and $1$ for $h\to h_{c2}$.  The spectrum (\ref{fse1}) allows to
identify the central charge $c=1$ and the operator content of a Gaussian
CFT with U(1) symmetry as mentioned above.

(ii) $\kappa={1\over4}$: proceeding as in (i) one finds $h_{c2}=3$. Unlike
for smaller $\kappa$ the dependence of the bare dispersion on the spectral
parameter near the minimum is $\propto \lambda^4$. This leads to a
different dependence of the magnetization on the field near $h_{c2}$:
\begin{equation}
  \sigma= 1-{4\over\pi} \left({3(h_{c2}-h)}\right)^{1\over4}\ .
\end{equation}
The classification of the low lying excitations as well as the
interpretation concerning the conformal properties of the system coincide
with those of case (i) above.

(iii) In the region ${1\over 4}<\kappa\le1$ two degenerate minima of the
bare dispersion exist at nonzero values $\lambda=\pm\Lambda^{(0)}$ of the
spectral parameter (Fig.~\ref{fig:phases}(b)).  This does not affect the
dependence of the magnetization on the magnetic field for fields near the
critical value
\begin{equation}
  h_{c2} = 1 + \kappa^{-{1\over2}}
\end{equation}
As for $\kappa<{1\over4}$ we have a square root dependence of $\sigma$ on
$h\lessapprox h_{c2}$
\begin{equation}
  \sigma = 1-\hbox{const.}\ \sqrt{h_{c2}-h}
\end{equation}
which a $\kappa$-dependent constant.
The spectrum of low energy excitations, however, is different in this
regime.  For $h<h_{c2}$ there are two filled ``Fermi seas''
$[-\Lambda_2,-\Lambda_1]$ and $[\Lambda_1,\Lambda_2]$ of quasiparticles
near $\pm\Lambda^{(0)}$ giving rise to two branches of massless spin
excitations with different magnon velocities in the system.  This situation
is very similar to the one observed in the system with XY-type anisotropy
and large staggered chiral field \cite{frahm:92}: the expressions for the
leading finite size corrections to the energies are (for the generalization
of \cite{vewo:85} to the case of several branches of excitations see also
\cite{izkr:89,woyn:89})
\begin{equation}
  \Delta E = -{\pi\over6L} \left( v_1+v_2 \right)
   + {2\pi\over L} \left( v_1(\Delta_1^+ + \Delta_1^-)
                        + v_2(\Delta_2^+ + \Delta_2^-) \right)\ .
\label{fse2}
\end{equation}
Here $v_i$ are the velocities of the excitations (linearized near the Fermi
points $\pm\Lambda_i$) and the primary conformal dimensions $\Delta_i^\pm$ are
found to be
\begin{equation}
  \Delta_i^\pm = {1\over2} \left\{ 
     {{\xi_{i}^+\Delta M^\pm - \xi_{i}^- \Delta M^\mp}\over
        {2((\xi_{i}^+)^2 - (\xi_{i}^-)^2)}}
     \mp (\xi_{i}^+ \Delta D^{\pm}+ \xi_{i}^- \Delta D^\mp)
     \right\}^2\ .
\label{dim2}
\end{equation}
The numbers $\Delta M^\pm$ are the difference between the number of quasi
particles in the Fermi seas in the ground state and excited state,
respectively.  $\Delta D^\pm$ counts the backscattering events in the
excited state (giving rise to excitations with momenta being multiples of
twice the Fermi momenta of the quasiparticles).  All conformal dimensions
can be parametrized by the four numbers $\xi_i^\pm=\xi^\pm(\Lambda_i)$
which are given in terms of the integral equations
\begin{eqnarray}
   &&\xi^+(\lambda) = 1 
        - \int_{\Lambda_1}^{\Lambda_2} K(\lambda-\mu) \xi^+(\mu)
        - \int_{\Lambda_1}^{\Lambda_2} K(\lambda+\mu) \xi^-(\mu)\ ,
\nonumber \\
   &&\xi^-(\lambda) =  
        - \int_{\Lambda_1}^{\Lambda_2} K(\lambda+\mu) \xi^+(\mu)
        - \int_{\Lambda_1}^{\Lambda_2} K(\lambda-\mu) \xi^-(\mu)\ .
\end{eqnarray}
(\ref{fse2}) is the generic form of a low energy spectrum in a system with
two different branches of massless excitations.  A finite size spectrum of
the form (\ref{fse2}) arises in many one dimensional systems, e.g.\
integrable spin chains \cite{izkr:89,fryu:90,frahm:92} and various models
of correlated electrons \cite{woyn:89,CorrEl} where the two critical degrees
of freedom are holon and spinon excitations.  In the present model it can be
interpreted as a realization of two $c=1$ Gaussian models.

Decreasing the magnetic field further the universality class changes again
at $h=h_{c1}$ where $\Lambda_{1}=0$ and $v_{1}$ vanishes leading to a
divergence of the low temperature specific heat $C\approx (\pi T/6)
\left(v_1^{-1}+v_2^{-1}\right)$.  At this point one of the massless modes
has a quadratic dispersion and the system undergoes a Pokrovsky--Talapov
transition into the U(1) Gaussian phase with central charge $c=1$ that was
already identified for $\kappa<{1\over4}$ above.  The value of $h_{c1}$ has
to be determined by numerical solution of the integral equations
(\ref{str:epsilon}).  The complete magnetic phase diagram is shown in
Fig.~\ref{fig:crit}.

\begin{figure}
\begin{center}
\leavevmode
\epsfxsize=0.8\textwidth
\epsfbox{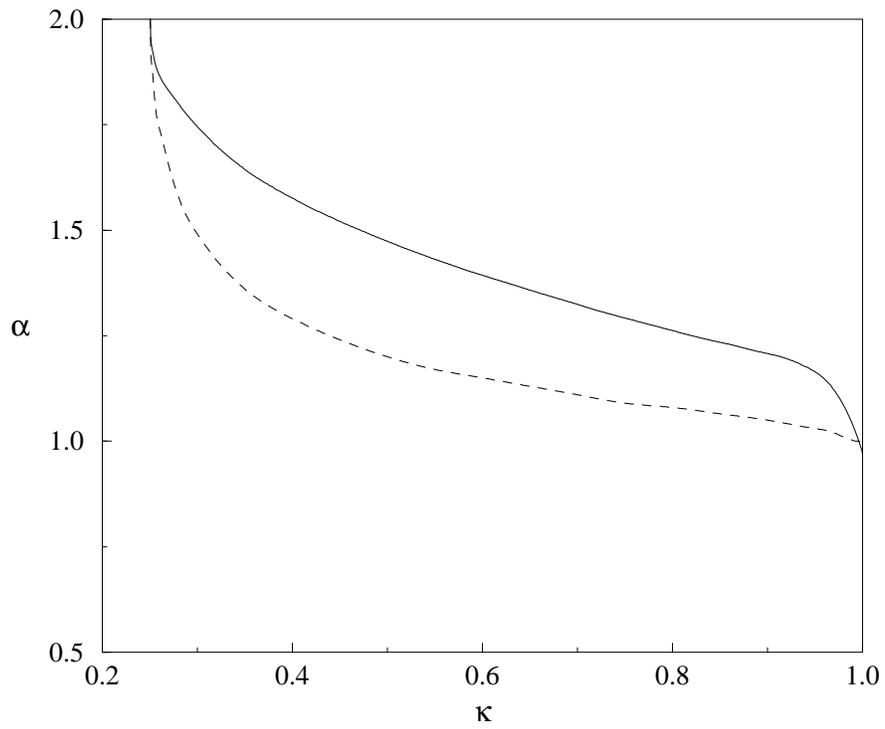}
\end{center}
\vspace*{.5cm}
\caption{
\label{fig:alpha} 
  Critical exponent $\alpha$ of $\langle S^z_x S^z_0 \rangle$ as a function
  of $\kappa$ for magnetic field $h\to h_{c1}$ from below (dashed line) and
  above (full line).}
\end{figure}

This phase transition leads to a discontinuous change in the spectrum of
critical exponents determining the long distance asymptotics of the
correlation functions of the system.  As an example we consider the spin spin
correlator $C^{zz}(x)=\langle S_x^{z}S_0^{z}\rangle$.  The massless magnon
excitations discussed above lead to algebraically decaying correlation
functions, from (\ref{dim1}) and (\ref{dim2}) we find that the leading term
in $C^{zz}$ beyond the constants is of the form $(1/x)^{\alpha}\
\cos(2k_{0}x)$ where $k_{0}$ is a magnetic field dependent wave number and the
exponent $\alpha$ is given by
\begin{equation}
  \alpha = \left\{ \begin{array}{cl}
      2\xi^2 & \mathrm{for~}0\le h<h_{c1}\\
      \sum_i \left((\xi^+_i)^2+(\xi^-_i)^2\right) & 
	\mathrm{for~}h_{c1}<h<h_{c2}\\
  \end{array} \right.\ .
\end{equation}
For $h\to0$ ($h_{c2}$) one obtains $\alpha=1$ ($2$) from the appropriate
expression as in region (i).  In Fig.~\ref{fig:alpha} the values of the
exponent $\alpha$ as $h_{c1}$ is approached from above and below are
presented as a function of $\kappa$.  For ${1\over4}<\kappa<1$ one observes
a discontinuous increase of $\alpha $ at the transition $h=h_{c1}$.  Such a
change of the asymptotic behaviour of $C^{zz}$ can be observed e.g.\ in the
temperature dependence of NMR longitudinal relaxation rate, $1/T_{1}\propto
T^{\alpha-1}$.

\begin{figure}
\begin{center}
\leavevmode
\epsfxsize=0.8\textwidth
\epsfbox{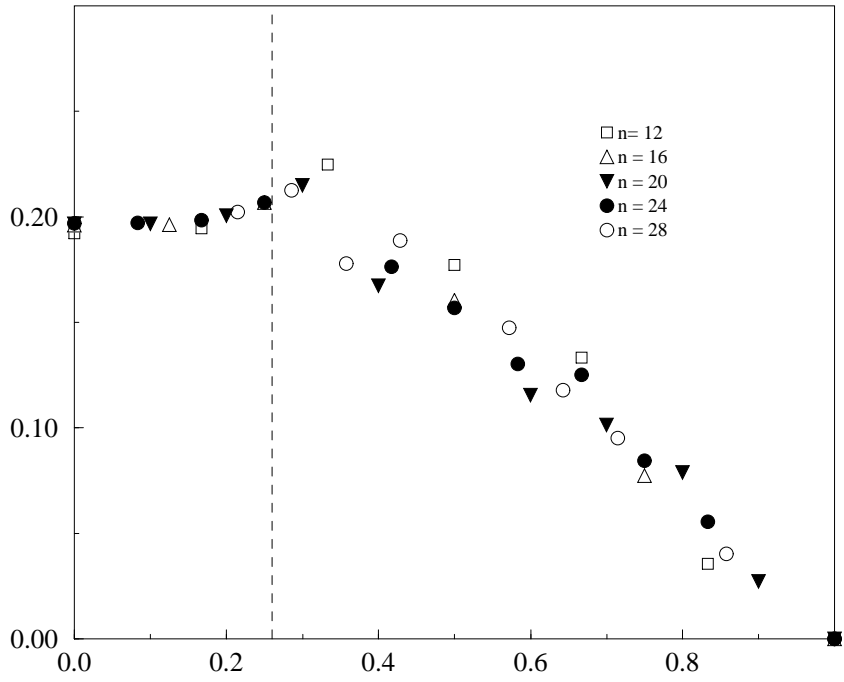}
\end{center}
\vspace*{.5cm}
\caption{
\label{fig:chi2} 
  Chiral fluctuations $\langle \left(\Delta\widehat{\chi}\right)^2 \rangle$
  in the ground state for given magnetization $\sigma$ of the integrable
  system with $\kappa={1\over2}$ and $n$ spins.  The dashed line indicates
  the value of the magnetization at the critical magnetic field $h_{c1}$
  for the infinite system.}
\end{figure}

Finally we want to note that that these pronounced features in the
$h$-dependence of the magnetization and the critical behaviour of the
system are reflected in the chirality properties of the system: Numerical
computation of the chirality in the ground state as a function of the
applied magnetic field for systems with up to 24 spins shows a smooth
dependence of $\langle\widehat{\chi}\rangle$ on the magnetization.  At the
same time the fluctuations $\langle \left(\Delta \widehat{\chi} \right)^2
\rangle$ scale like $1/N$ for $h<h_{c1}$ and are enhanced as the critical
field is approached from below.  Above $h_{c1}$ they become smaller
vanishing for the ferromagnetic state (see Figure~\ref{fig:chi2}).

\section{Summary and conclusion}
We have present results on the ground state properties and the magnetic
phase diagram of a quasi one-dimensional spin-system with explicitly $P$
and $T$ breaking terms in the Hamiltonian.  These terms generate chiral
order in the system which in turn is found to lead to an interesting
behaviour of the system when exposed to an external magnetic field.

\begin{figure}
\begin{center}
\leavevmode
\epsfxsize=0.8\textwidth
\epsfbox{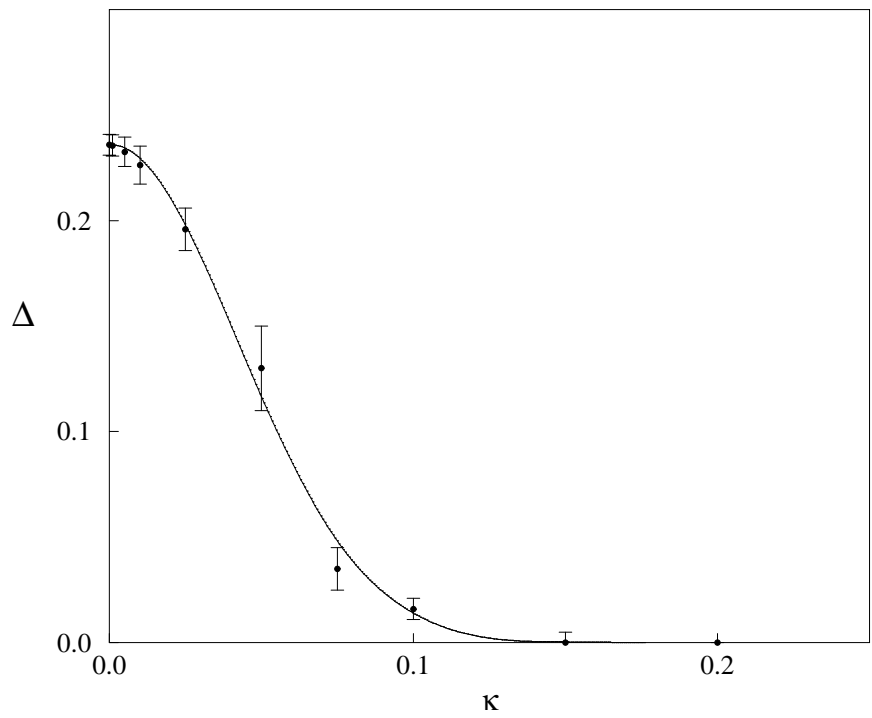}
\end{center}
\vspace*{.5cm}
\caption{
\label{fig:gap} 
  Spin gap $\Delta$ for $J_1=2J_2=1$ as a function of the strength of the
  chiral field. The line is guide for the eyes and is given by a Gaussian.}
\end{figure}

The question remains to which extent the observed behaviour is determined
by the integrability of the system (\ref{hami}), (\ref{parint}).  While a
complete analyisis of this question is difficult we have diagonalized small
systems to give a partial answer:
 
Choosing the parameters as in (\ref{choice}) with $J_2={1\over2}$ the
system interpolates between two soluble points: For $\kappa=0$ the model
reduces to the MG--model, $\kappa=\frac{1}{2}$ is the Bethe Ansatz soluble
point.  While the former one has a gap for spin excitations, the latter
supports massless magnons.  The vanishing of this spin gap as a function of
$\kappa$ has been studied using Lanzcos procedures and finite size
interpolation. At the MG--point the gap above the two degenerate valence
bond singlets is known very accurately, being $\Delta_{MG} = 0.236$ with
finite size corrections scaling as $\Delta^2=\Delta_{MG}^2(1+
\mathrm{const.}/{N^2})$ \cite{MGGap}.  At $\kappa=\frac{1}{2}$ the gap
vanishes as $\Delta \sim \mathrm{const}/{N}$ according to (\ref{dim1}).  It
is difficult to find an interpolating expression between these limiting
cases that leads to uniformly good fits of the numerical data.  Still one
can conclude that inclusion of the chiral term results in a rapidly
vanishing gap, as can be seen in Figure \ref{fig:gap}.  For
$\kappa\gtrapprox0.15$ the numerical data do not allow to decide whether
there is a finite gap or not.
The ground state expectation value of the chirality shows a monotonic
increase with $\kappa$, with $\langle\widehat{\chi}\rangle \approx 0.58$ at
$\kappa=1$.

Further studies are necessary for a better understanding of the
intermediate phase transition in the integrable model at $h=h_{c1}$.  At
$h=0$ this transition occurs at the point $\kappa=1$, corresponding to a
decoupling of the two sublattices.  Whether such a `dimensional reduction'
is the origin of the phase transition at $\kappa<1$ and whether this
feature can be used to describe the experimentally observed magnetic phases
in the frustrated ABX$_3$ compounds remains to be investigated in the
context of systems with a larger number of coupled chains and eventually
truly two-dimensional lattices.

\section*{Acknowledgements}
We gratefully acknowledge useful discussions with H.-U.\ Everts, C.\
Lhuillier, U. Neugebauer, C. Waldtmann and A.~A.\ Zvyagin.  This work has
been supported by the Deutsche Forschungsgemeinschaft under Grant No.\
Fr~737/2--2.  The numerical calculations have been performed partly at the
Regionales Rechenzentrum f\"ur Niedersachsen, Hannover, and the Zuse
Rechenzentrum, Berlin.


\newpage

\end{document}